\documentclass[aps,10pt,pre,twocolumn,superscriptaddress,floatfix]{revtex4-2}
\usepackage{graphicx}
\usepackage{xcolor}
\usepackage{subfigure,amsmath, amsthm, amssymb}
\usepackage[hidelinks]{hyperref}
\usepackage{csquotes}
\graphicspath{{./Figures/}}
\usepackage[normalem]{ulem} 

\newcommand{\be}{\begin{equation}}
\newcommand{\ee}{\end{equation}}
\newcommand{\bea}{\begin{eqnarray}}
\newcommand{\eea}{\end{eqnarray}}
\begin{document}

\title{Slow heat-driven flow in a gas of hard disks}

\author{Amit Kumar}
\email{amit.k@icts.res.in}
\affiliation{International Centre for Theoretical Sciences, Tata Institute of Fundamental Research, Bangalore 560089, India}
\author{Abhishek Dhar}
\email{abhishek.dhar@icts.res.in}
\affiliation{International Centre for Theoretical Sciences, Tata Institute of Fundamental Research, Bangalore 560089, India}

\author{Baruch Meerson}
\email{meerson@mail.huji.ac.il}
\affiliation{Racah Institute of Physics, Hebrew University of
Jerusalem, Jerusalem 91904, Israel}

\begin{abstract}

We study a slow heat-driven flow in a gas of elastically colliding hard disks
confined to a long channel. The initial state consists of two regions with
large temperature and density contrasts but nearly equal pressures, leading to
a low-Mach-number, nearly isobaric evolution. In the dilute limit, the
corresponding isobaric hydrodynamic theory reduces to a previously known
ideal-gas description. We extend this theory to finite densities by
incorporating a non-ideal equation of state of a hard-disk fluid, and solve
the resulting one-dimensional equations numerically. Finite-density effects
produce appreciable deviations from the ideal-gas prediction. We then test the
theory directly against event-driven molecular dynamics simulations of hard
disks and find very good agreement in both the dilute and finite-density
regimes. The results provide, to our knowledge, the first particle-level test
of isobaric gas dynamics of a strongly inhomogeneous cooling
flow.
\end{abstract}

\maketitle

\section{Introduction}

Connecting continuum hydrodynamics quantitatively to the underlying microscopic dynamics remains an important problem in nonequilibrium statistical mechanics. Direct numerical tests provide one route to addressing this connection. In recent years, molecular dynamics simulations of hard-particle systems have enabled detailed comparisons between microscopic dynamics and hydrodynamic predictions in strongly compressible, rapidly evolving flows. In particular, quantitative agreement has been demonstrated in strongly nonequilibrium situations involving shocks and blast waves \cite{joy2021shock3d,joy2021shock,chakraborti2021blast,ganapa2021blast,chakraborti2022splash,kumar2022blast,singh2023blast,kumar2024shock,kumar2025shock,kumar2025splashinhomogeneousgasdimension,kumar2025impurity}. These studies have also revealed regimes in which hydrodynamic and non-hydrodynamic structures coexist, thereby illustrating both the predictive power and the limitations of continuum descriptions.

A complementary situation arises when the characteristic evolution time is long compared with the acoustic time. Sound waves then rapidly establish mechanical equilibrium, while temperature, density, and velocity continue to evolve on a much longer time scale. In the absence of external forces this corresponds to an approximately uniform pressure, whereas in the presence of gravity it corresponds to an approximately hydrostatic state. The resulting asymptotic reduction filters out acoustic dynamics and is commonly referred to as the low-Mach-number or zero-Mach-number approximation. It has been developed in a variety of contexts, including thermally driven flows, combustion, and atmospheric dynamics \cite{RehmBaum1978,MajdaSethian1985,Majda1984,Williams1985,PoinsotVeynante2005,OguraPhillips1962,Durran1989}, as well as slow flows of nearly elastic granular gases \cite{Goldhirsch2003,BrombergLivneMeerson2003,VolfsonMeersonTsimring2006}. Importantly, a small hydrodynamic velocity compared with the speed of sound does not imply small variations of temperature or density.

An important class of problems within this framework is the slow relaxation of strongly nonuniform temperature fields. In the absence of external forces, heat transport is accompanied by mass redistribution while the pressure remains nearly uniform. Earlier work has shown that strongly heated localized regions can exhibit nontrivial cooling dynamics and self-similar behavior \cite{Meerson1989,Kaganovich,GLM,MSS,Kamin}.

Compared with the high-Mach-number situations discussed above, direct microscopic tests of slow, thermally driven low-Mach-number dynamics appear to be lacking. In this regime, the rapid acoustic degrees of freedom have effectively relaxed, and the subsequent evolution is controlled by heat conduction and the associated slow redistribution of mass. A comparison between the particle dynamics and the reduced isobaric hydrodynamic description therefore provides a complementary test of continuum hydrodynamics in a regime very different from shocks and blast waves.

In this work, we perform such a comparison for a gas of elastically colliding hard disks in a long channel, where the macroscopic flow is effectively one-dimensional. The initial state consists of two regions with markedly different temperatures and densities but nearly equal pressures, leading to a strongly nonlinear yet nearly isobaric evolution. In the dilute-gas limit, the corresponding hydrodynamic description reduces to the previously known ideal-gas theory of Ref.~\cite{Meerson1989}. We extend this description to finite densities by incorporating the non-ideal equation of state and the corresponding transport properties of a hard-disk fluid. We solve the resulting hydrodynamic equations numerically and compare their predictions with event-driven molecular dynamics simulations. To our knowledge, this provides the first direct microscopic test of zero-Mach-number hydrodynamics, both in the dilute-gas limit and at finite density.

The remainder of the paper is organized as follows. In Sec.~\ref{eqs} we summarize the formalism \cite{Meerson1989} for the isobaric heat-driven flow of an ideal gas and specialize it to the quasi-one-dimensional setting described above. In Sec.~\ref{finite_density_form} we extend the hydrodynamic description to a finite-density gas of hard disks, introduce dimensionless variables, and derive the corresponding self-similarity equations. In Sec.~\ref{HDsolutions} we analyze numerical solutions of these equations and the effects of finite density. In Sec.~\ref{MD} we describe the event-driven molecular dynamics simulations and compare them with the hydrodynamic predictions. Section~\ref{Discussion} summarizes and discusses our results.

\section{Isobaric heat-driven flow of ideal gas}
\label{eqs}
Consider a long channel filled with ideal gas.  The
initial temperature and density of the gas is $T_L$ and $\rho_L$ at $x<0$,
$T_R$ and $\rho_R$ at $x>0$. As we want
to study a zero-Mach-number heat-conduction process, we choose the
parameters $T_L,T_R,\rho_L$ and $\rho_R$ to obey the relation
\begin{equation}\label{isobaric0}
\rho_L T_L = \rho_R T_R,
\end{equation}
so that the ideal gas pressure is constant through the system. 

We will follow the evolution of the temperature and density of the gas over
times much longer than the typical sound travel time 
through the system. The basic process here is heat transfer from the hot
part of the gas to the cold part. Importantly, the gas density also changes: 
there is a slow flow of the
gas from the dense (cold) to dilute (hot) region so that the
pressure over the system remains approximately constant. This implies
that the density profile is ``enslaved" by the temperature profile (or
vice versa), so that
\begin{equation}\label{isobaric}
\rho(x,t) T(x,t)  \simeq \text{const.}
\end{equation}
at all times.

The governing equation for the temperature is a planar analog of the nonlinear heat equation \cite{Meerson1989}
originally derived for a study of the isobaric cooling flow which develops after a strong explosion in a gas. 
In the present quasi-one-dimensional setting the equation has the form
\begin{equation}\label{pde}
  \partial_t T  =\beta_0 T^2 \partial_x \left(T^{-1/2} \partial_x T\right)\,.
\end{equation}
The dimensional coefficient $\beta_0$ is defined by the relation
\begin{align}
  \beta(T)\equiv\frac{\kappa(T)}{c_p \rho_R T_R} \equiv \beta_0 T^{1/2},\label{betaDef}  
\end{align}
where $\kappa(T) \sim T^{1/2}$ and $c_p$ are the heat conductivity and
the isobaric heat capacity of ideal gas, respectively.

Once the temperature $T(x,t)$ is determined, one can calculate the 
gas density  $\rho(x,t) = \rho_R T_R/T(x,t)$ and the heat-driven flow 
velocity \cite{Meerson1989}
\begin{equation}\label{vxt}
 v(x,t) = \beta_0 T^{1/2} \partial_x T\,.
\end{equation}
In our quasi-one-dimensional setting Eq.~(\ref{pde}) should be solved subject to the boundary conditions 
\begin{equation}\label{BCs}
T(x=-\infty,t)=T_L \quad \text{and} \quad T(x=\infty,t)=T_R
\end{equation}
and the initial condition 
\begin{equation}\label{IC}
T(x,0)=T_L+(T_R-T_L) \,\theta (x)\,,
\end{equation}
where $\theta(x)$ is the Heaviside step-function. 

\begin{figure}[ht]
\includegraphics[width=0.48\textwidth]{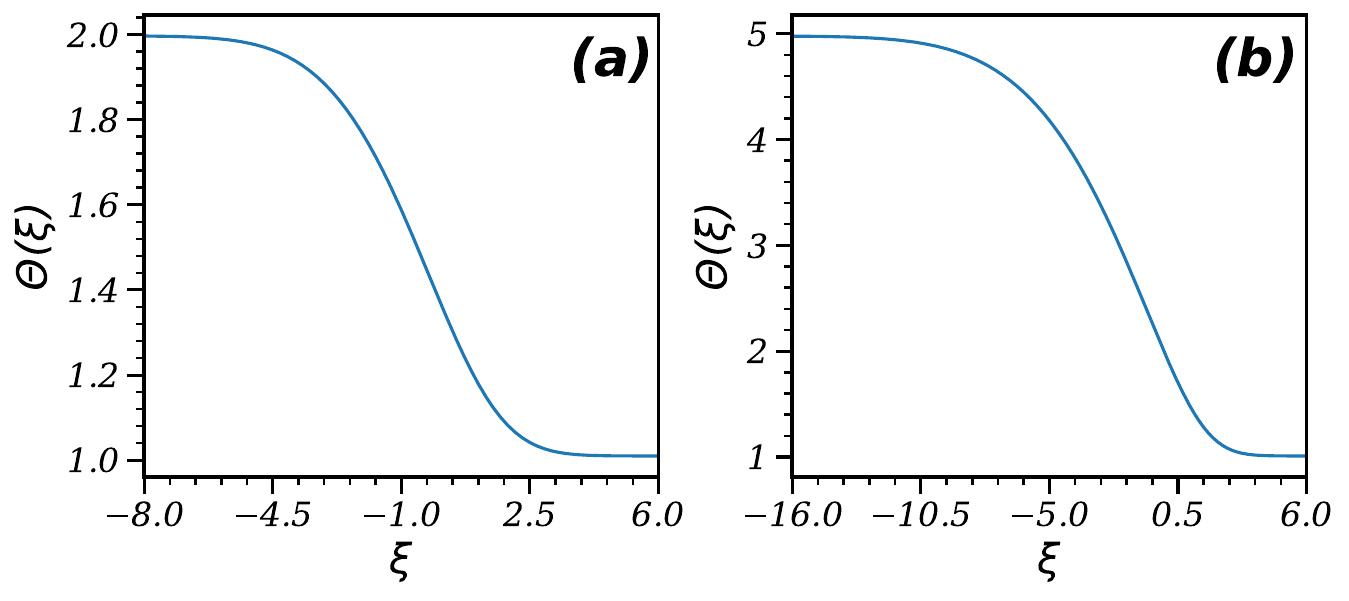}
\caption{Scaling function $\Theta(\xi)$ for $T_L/T_R =2$ (left panel) and $5$ (right panel).}
\label{Thetanum}
\end{figure}

Since the formulation of this problem does not include any quantity with the dimension of length,
the solution is self-similar starting from $t=0$, and the similarity ansatz immediately follows from dimensional analysis \cite{Barenblatt}. The dimensions of
$\beta_0$ can be found from Eq. (\ref{pde}):
\begin{equation}\label{beta0dim}
[\beta_0]=\frac{\text{length}^2}{\text{time} \times \text{temperature}^{3/2}}\,.
\end{equation}
This leads to the similarity ansatz 
\begin{equation}
T(x,t)=T_R \,\Theta\left[\frac{x}{T_R^{3/4}\,(\beta_0 t)^{1/2}}\right]\,.\label{rescTideal}
\end{equation} 
Plugging it into Eq.~(\ref{pde}), we obtain an ordinary differential equation (ODE) for the scaling function $\Theta(\xi)$:
\begin{equation}\label{ODE}
\Theta^2\,\frac{d}{d\xi}\left(\Theta^{-1/2}\,\frac{d\Theta}{d\xi}\right)+\frac{\xi}{2}\, \frac{d\Theta}{d\xi}=0\,.
\end{equation}
This ODE should be solved subject to the boundary conditions $\Theta(-\infty)=T_L/T_R$ and $\Theta(\infty)=1$. 
For a given $T_L/T_R$ the solution for the scaling function $\Theta(\xi)$ can be obtained numerically by the shooting method~\cite{shooting}. Two examples of numerically found $\Theta(\xi)$ are shown in Fig. \ref{Thetanum}.

These hydrodynamic calculations can be applied, in particular, to a dilute gas of hard disks. The bulk thermal conductivity $\kappa(T)$ of such a gas does exhibit a square-root dependence on the temperature, as assumed above ~\cite{huang1963statistical,reif2009fundamentals}:
\begin{align}
\kappa = C_{\rm dil} \sqrt{T}. \label{kappaDefKTG}
\end{align}
The coefficient $C_{\rm dil}$ is known from the Chapman-Enskog expansion~\cite{dorfman2021contemporary,gass1971enskog,garcia2006transport,kremer2010introduction} and at the lowest order is independent of $\rho$: 
$C_{\rm dil} = {2/\sqrt{\pi}}$, for hard disks of unit diameter.

The picture is a bit more complicated in two dimensions,  where hydrodynamic current correlations decay at long times as \(1/t\), so Green--Kubo integrals for transport coefficients, including thermal conductivity, acquire a logarithmic infrared/system-size dependence
\cite{AlderWainwright1970,ErnstHaugeLeeuwen1970,PomeauResibois1975}. However, the divergence is weak, scaling logarithmically with system size. In this work we use the finite bare transport coefficient given by kinetic theory with a dimensionless correction factor, $C_1$, to account for possible finite-size effects. This constant is determined  by comparison with the event-driven molecular dynamics (EDMD) simulations.

\section{\label{finite_density_form} Isobaric heat-driven flow at finite density}

We now extend the description to a finite-density gas of hard disks. As before, the two halves of the long channel are initially prepared at different temperatures and number densities, with equal pressures. To make the nondimensionalization explicit, we temporarily append a subscript ${\rm d}$ to all dimensional fields and far-field quantities. After the change of variables below, quantities without this subscript are dimensionless.

Neglecting viscous stresses, the one-dimensional hydrodynamic equations for the number density $\rho_{\rm d}(x_{\rm d},t_{\rm d})$, longitudinal velocity $v_{\rm d}(x_{\rm d},t_{\rm d})$, temperature $T_{\rm d}(x_{\rm d},t_{\rm d})$, and pressure $p_{\rm d}(x_{\rm d},t_{\rm d})$ are given by~\cite{landaubook,warsi2005fluid}
\begin{subequations}\label{eq:hydro_dim}
\begin{align}
& \partial_{t_{\rm d}} \rho_{\rm d}+\partial_{x_{\rm d}}(\rho_{\rm d}v_{\rm d})=0,\label{eq:hydroa_dim}\\
& m\rho_{\rm d}\left(\partial_{t_{\rm d}}v_{\rm d}+v_{\rm d}\partial_{x_{\rm d}}v_{\rm d}\right)+\partial_{x_{\rm d}}p_{\rm d}=0,\label{eq:hydrob_dim}\\
& \rho_{\rm d}c_v\left(\partial_{t_{\rm d}}T_{\rm d}+v_{\rm d}\partial_{x_{\rm d}}T_{\rm d}\right)
+p_{\rm d}\partial_{x_{\rm d}}v_{\rm d}
=\partial_{x_{\rm d}}\!\left(\kappa_{\rm d}\partial_{x_{\rm d}}T_{\rm d}\right),\label{eq:hydroc_dim}
\end{align}
\end{subequations}
where $c_v=k_B/(\gamma-1)$ is the heat capacity at constant volume per particle. For a two-dimensional monatomic gas $\gamma=2$.

We use the disk diameter $d$ as the unit of length and introduce an arbitrary fixed reference temperature $T_*$. The corresponding thermal velocity and microscopic time are
\begin{equation}
 v_* = \sqrt{\frac{k_BT_*}{m}},\qquad
 t_* = \frac{d}{v_*}=d\sqrt{\frac{m}{k_BT_*}}.
\end{equation}
The complete change of variables is
\begin{equation}\label{eq:nondimdefs}
\begin{gathered}
 x_{\rm d}=dx,\; y_{\rm d}=dy,\; t_{\rm d}=t_*t,\\
 \rho_{\rm d}=d^{-2}\rho,\; T_{\rm d}=T_*T,\; v_{\rm d}=v_*v,\\
 p_{\rm d}=\frac{k_BT_*}{d^2}p,\qquad
 \kappa_{\rm d}=\frac{k_Bv_*}{d}\kappa.
\end{gathered}
\end{equation}
The channel dimensions and far-field quantities are therefore rescaled according to
\begin{equation}\label{eq:farfieldscales}
\begin{gathered}
 L_x=\frac{L_{x,{\rm d}}}{d},\qquad
 L_y=\frac{L_{y,{\rm d}}}{d},\\
 \rho_L=\rho_{L,{\rm d}}d^2,\qquad
 \rho_R=\rho_{R,{\rm d}}d^2,\\
 T_L=\frac{T_{L,{\rm d}}}{T_*},\qquad
 T_R=\frac{T_{R,{\rm d}}}{T_*},\qquad
 p_0=\frac{p_{0,{\rm d}}d^2}{k_BT_*}.
\end{gathered}
\end{equation}
Thus $T_*$ only fixes the units; it is independent of the particular left and right states.

Substituting Eq.~\eqref{eq:nondimdefs} into Eqs.~\eqref{eq:hydro_dim} gives the fully dimensionless hydrodynamic equations
\begin{subequations}
\begin{align}
& \partial_t\rho+\partial_x(\rho v)=0,\label{eq:hydroa}\\
& \rho\left(\partial_t v+v\partial_xv\right)+\partial_xp=0,\label{eq:hydrob}\\
& \frac{\rho}{\gamma-1}\left(\partial_tT+v\partial_xT\right)
+p\,\partial_xv
=\partial_x\!\left(\kappa\,\partial_xT\right).\label{eq:hydroc}
\end{align}
\end{subequations}
All dependence on $m$, $d$, $k_B$, and the chosen reference scale $T_*$ has been absorbed into the units in Eq.~\eqref{eq:nondimdefs}.

For the equation of state we use
\begin{equation}
 p=\rho T Z(\rho),\label{eq:HS-EOS}
\end{equation}
where the packing fraction is
\begin{equation}
 \nu(\rho)=\frac{\pi\rho}{4},
\end{equation}
and the compressibility factor is approximated by the Henderson form~\cite{henderson1975simple,barrat2002molecular,FiniteWidthEOSNote},
\begin{equation}
 Z(\rho)=\frac{1+\nu(\rho)^2/8}{[1-\nu(\rho)]^2}
 =\frac{128+\pi^2\rho^2}{8(4-\pi\rho)^2}.
\end{equation}

For the finite-density thermal conductivity we use the kinetic-theory expression~\cite{jenkins1985kinetic}. It is useful to separate its known density and temperature dependence from an overall dimensionless correction factor. We write
\begin{equation}\label{kappaDefOrig}
 \mathcal K(\rho,T)=
 \frac{2\rho\,G(\rho)\sqrt{T}}{\sqrt{\pi}}
 \left[1+\frac{9\pi}{16}\left(1+\frac{2}{3G(\rho)}\right)^2\right],
\end{equation}
where
\begin{equation}
 G(\rho)=
 \frac{\nu(\rho)[16-7\nu(\rho)]}
 {16[1-\nu(\rho)]^2},
\end{equation}
and take
\begin{equation}
 \kappa(\rho,T)=C_1\,\mathcal K(\rho,T).\label{kappaDef}
\end{equation}
Here $C_1$ is an adjustable dimensionless constant $O(1)$, which will be determined by comparison with the molecular-dynamics simulations. It should not be confused with the coefficient $C_{\rm dil}$ in the dilute-gas expression~\eqref{kappaDefKTG}. In the dilute limit,
\begin{equation}
 \mathcal K(\rho,T)\simeq \frac{2}{\sqrt\pi}\sqrt{T},\qquad \rho\to0,
\end{equation}
so that $C_1=1$ corresponds to the lowest-order dilute hard-disk result.

The dimensionless far-field conditions are
\begin{subequations}\label{farfield}
\begin{align}
& T(x\to-\infty,t)=T_L,
\qquad T(x\to\infty,t)=T_R,\label{farfield:a}\\
& \rho(x\to-\infty,t)=\rho_L,
\qquad \rho(x\to\infty,t)=\rho_R,\label{farfield:b}\\
& v(x\to-\infty,t)=0,
\qquad v(x\to\infty,t)=0.\label{farfield:c}
\end{align}
\end{subequations}
The equal-pressure preparation imposes
\begin{equation}\label{eq:equalpressuredimless}
 p_0=\rho_LT_L Z(\rho_L)=\rho_RT_R Z(\rho_R),
\end{equation}
so that, for example,
\begin{equation}\label{eq:TLTRp0}
 T_R=\frac{p_0}{\rho_RZ(\rho_R)},\qquad
 T_L=\frac{p_0}{\rho_LZ(\rho_L)}.
\end{equation}
A convenient set of independent dimensionless parameters is therefore $L_x$, $L_y$, $\rho_L$, $\rho_R$, $T_R$ (or, equivalently, $p_0$), $\gamma$, and $C_1$. The ratio $T_L/T_R$ is fixed by Eq.~\eqref{eq:equalpressuredimless}. In the effectively infinite one-dimensional problem $L_x$ and $L_y$ do not enter the reduced equations explicitly; they control finite-size effects and the validity of the quasi-one-dimensional approximation.

In the low-Mach-number regime the momentum equation~\eqref{eq:hydrob} can be replaced, to leading order, by the isobaricity condition
\begin{equation}
 p=\rho T Z(\rho)\simeq p_0=\mathrm{const}.\label{eq:isobar}
\end{equation}
The remaining equations are Eqs.~\eqref{eq:hydroa}, \eqref{eq:hydroc}, and \eqref{eq:isobar}.

In the dilute limit, Eq.~\eqref{kappaDef} gives $\kappa\simeq(2C_1/\sqrt{\pi})\sqrt{T}$. Since $c_p=\gamma/(\gamma-1)$ in the present dimensionless units, we introduce the convenient dimensionless diffusive scale
\begin{equation}\label{eq:Dtilde}
 D=\frac{2C_1\sqrt{T_R}}{\sqrt{\pi}\,\rho_R}
 \frac{\gamma-1}{\gamma},
\end{equation}
whose dimensional counterpart,
\begin{equation}\label{eq:Ddim}
 D_{\rm d}=dv_*D
 =\frac{2C_1}{\sqrt{\pi}\,d\,\rho_{R,{\rm d}}}
 \frac{\gamma-1}{\gamma}
 \sqrt{\frac{k_BT_{R,{\rm d}}}{m}},
\end{equation}
is independent of the arbitrary reference temperature $T_*$.

We next exploit the scaling symmetry of the isobaric finite-density hydrodynamic equations. The step profile at $t=0$ and the time-independent far-field states are invariant under
\begin{equation}\label{eq:scalingsymmetry}
 x\to ax,\qquad t\to a^2t,\qquad v\to a^{-1}v,
\end{equation}
with $\rho$, $T$, and $p$ unchanged. It is this invariance, together with the fixed far-field amplitudes, rather than merely the absence of an initial length scale, that leads to diffusive self-similarity.

We choose the similarity variable and scaling functions as
\begin{equation}\label{scaledForm}
\begin{gathered}
 \xi=\frac{x}{\sqrt{Dt}}
 =\frac{x_{\rm d}}{\sqrt{D_{\rm d}t_{\rm d}}},\\
 \frac{\rho}{\rho_R}=R(\xi),\qquad
 \frac{T}{T_R}=\Theta(\xi),\\
 v=\sqrt{\frac{D}{t}}\,V(\xi).
\end{gathered}
\end{equation}
Because $\mathcal K(\rho,T)$ is proportional to $\sqrt{T}$, one has $\mathcal K(\rho,T_R\Theta)=\sqrt{T_R}\,\mathcal K(\rho,\Theta)$. With this normalization, both $T_R$ and the overall factor $C_1$ drop out of the equations for the scaling functions. Substitution of Eq.~\eqref{scaledForm} into Eqs.~\eqref{eq:hydroa} and \eqref{eq:hydroc} gives 
\begin{align}
&\left(V-\frac{\xi}{2}\right)\frac{dR}{d\xi}+R\frac{dV}{d\xi}=0,\label{contMassReducedHardC_ND}\\
&R\left(V-\frac{\xi}{2}\right)\frac{d\Theta}{d\xi}
+\overline p_0\frac{dV}{d\xi}
=\frac{\gamma\sqrt \pi}{2}\frac{d}{d\xi}
\left[\mathcal K(\rho_RR,\Theta)\frac{d\Theta}{d\xi}\right],\label{contEneReducedHardC_ND}
\end{align}
where
\begin{equation}
 \overline p_0=(\gamma-1)\frac{p_0}{\rho_RT_R}
 =(\gamma-1)Z(\rho_R).
\end{equation}
The isobaricity condition becomes
\begin{equation}
 p_0=\rho_RT_R R(\xi)\Theta(\xi)
 Z[\rho_RR(\xi)],\label{p0scaled}
\end{equation}
or, equivalently,
\begin{equation}\label{eq:ThetaofR}
 \Theta(\xi)=\frac{Z(\rho_R)}
 {R(\xi)Z[\rho_RR(\xi)]}.
\end{equation}
Equations~\eqref{contMassReducedHardC_ND}, \eqref{contEneReducedHardC_ND}, and \eqref{eq:ThetaofR} reduce the problem to two coupled ODEs for $R$ and $V$. We solve them with
\begin{equation}
 R(\infty)=1,\qquad
 R(-\infty)=\frac{\rho_L}{\rho_R},\qquad
 V(\infty)=0,\label{bc:a}
\end{equation}
using a shooting method~\cite{shooting}. The remaining far-field conditions then follow from the equations and the isobaric relation.

Two consequences of this rescaling are worth emphasizing. First, the scaling functions $R(\xi)$, $\Theta(\xi)$, and $V(\xi)$ are independent of the overall conductivity factor $C_1$; this factor fixes only the conversion between $\xi$ and $x$ through $D$. Second, at fixed $\rho_L$ and $\rho_R$, changing the common pressure $p_0$ changes $T_L$, $T_R$, and therefore $D$, but not the scaling functions. Thus the values of $p_0$, $T_L$, and $T_R$ used below can be kept in the same dimensionless units as in the molecular-dynamics simulations.

In the dilute limit $\rho_L,\rho_R\to0$ at fixed density ratio, $Z\to1$ and $\mathcal K\to2\sqrt{\Theta/\pi}$. Equations~\eqref{contMassReducedHardC_ND}--\eqref{eq:ThetaofR} then give
\begin{align}
    &R=\Theta^{-1},\label{idealRxi}\\
    &V=\Theta^{1/2}\frac{d\Theta}{d\xi}.\label{idealVxi}
\end{align}
Eliminating $R$ and $V$ reproduces Eq.~\eqref{ODE} of the ideal-gas theory with the same similarity coordinate $\xi$.

\section{Finite-Density Solutions}
\label{HDsolutions}
For the numerical illustrations below we use the same dimensionless units that will be employed in the EDMD simulations, namely $d=m=k_B=T_*=1$. Thus the numerical values of $p_0$, $T_L$, and $T_R$ are the dimensionless quantities defined in Sec.~\ref{finite_density_form}.

\begin{figure*}[t]
\centering
\includegraphics[width=0.8\textwidth]{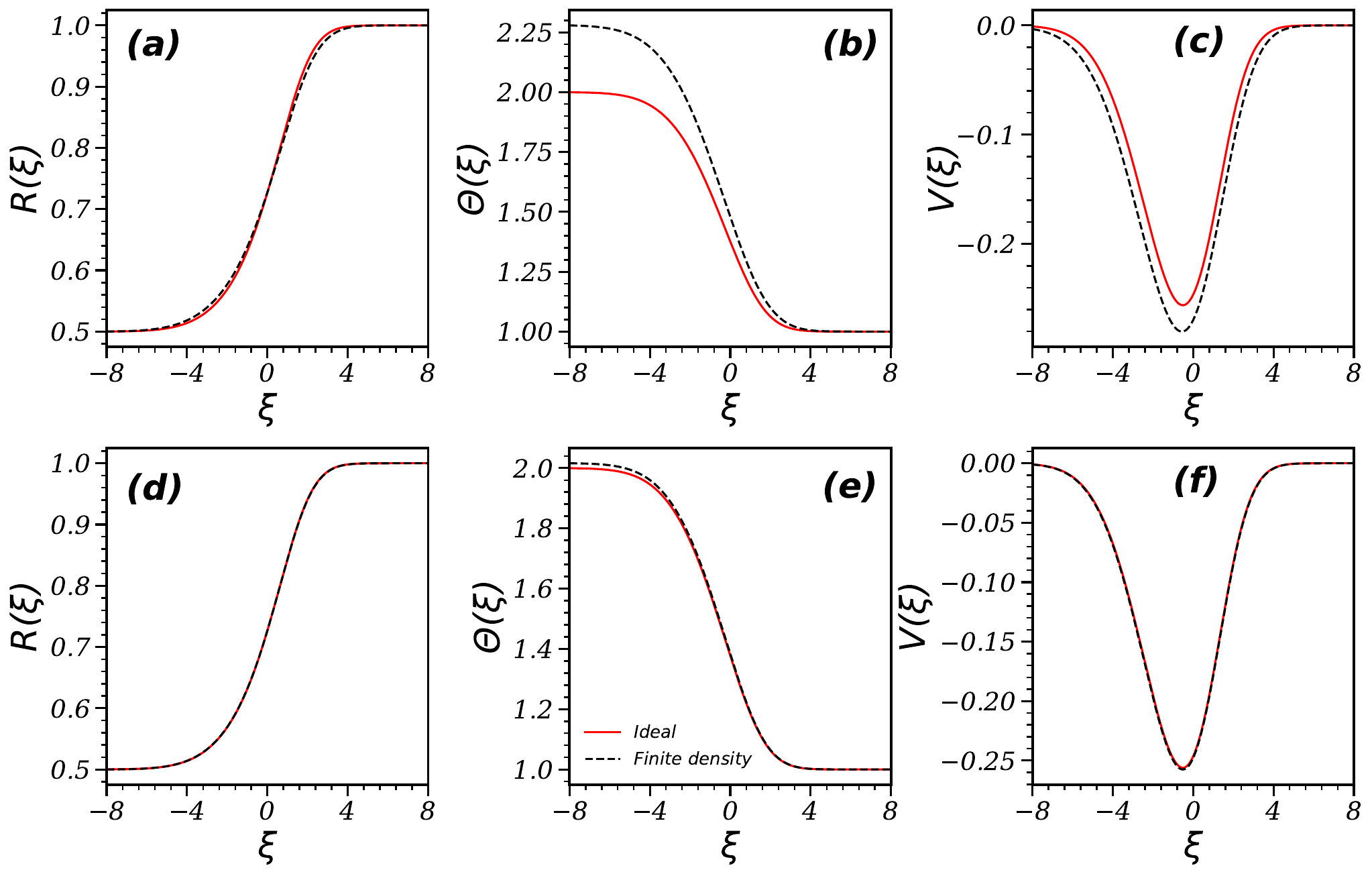}
\caption{Finite-density hard-disk gas versus the ideal-gas limit. Shown are the rescaled density $R$, temperature $\Theta$, and velocity $V$ as functions of the similarity coordinate $\xi$. Panels (a--c) correspond to $\rho_L=0.075$ and $\rho_R=0.15$, while panels (d--f) correspond to $\rho_L=0.005$ and $\rho_R=0.01$. The finite-density curves are obtained by solving the isobaric finite-density similarity equations, while the ideal-gas curves are obtained from Eq.~\eqref{ODE} together with Eqs.~\eqref{idealRxi} and \eqref{idealVxi}. As the densities decrease, the finite-density profiles converge to their ideal-gas counterparts. In each model the left and right temperatures are fixed by the equal-pressure condition with the corresponding equation of state; for the finite-density curves, $p_0=10$, $T_R=p_0/[\rho_R Z(\rho_R)]$, and $T_L=p_0/[\rho_L Z(\rho_L)]$.}
\label{compare_p0_ideal_hard_compare_0.005_0.01}
\end{figure*}

We first quantify the finite-density corrections by comparing the numerical solutions of the isobaric finite-density similarity equations with the ideal-gas solution of Sec.~\ref{eqs}, obtained from Eq.~\eqref{ODE} together with Eqs.~\eqref{idealRxi} and \eqref{idealVxi}. In the dilute limit, $(\rho_L,\rho_R)\to(0,0)$ at fixed density ratio, the hard-disk equation of state and transport coefficients reduce to their ideal-gas forms, so the two solutions must converge. Figure~\ref{compare_p0_ideal_hard_compare_0.005_0.01} shows this comparison for $(\rho_L,\rho_R)=(0.075,0.15)$ [panels (a--c)] and $(0.005,0.01)$ [panels (d--f)]. At the larger densities the finite-density corrections are clearly visible, whereas at the smaller densities the two sets of profiles are already very close.

Two further properties of the similarity solutions will be tested against the EDMD simulations in Sec.~\ref{MD}. First, at fixed $\rho_L$ and $\rho_R$, changing the common pressure $p_0$ changes $T_L$, $T_R$, and hence the spatial scale through $D$, but leaves the scaling functions $R(\xi)$, $\Theta(\xi)$, and $V(\xi)$ unchanged. Second, at fixed $\rho_R$ the scaling functions do change when the density ratio $\rho_L/\rho_R$ is varied.

As shown in Sec.~\ref{finite_density_form}, $C_1$ enters only through the diffusive scale $D$ and therefore drops out of the equations for the scaling functions. It affects the profiles in physical coordinates through the conversion $\xi=x/\sqrt{Dt}$, but not the functions $R(\xi)$, $\Theta(\xi)$, and $V(\xi)$ themselves.

\section{Event-Driven Molecular Dynamics and Comparison with Hydrodynamics}
\label{MD}

We now describe the event-driven molecular dynamics (EDMD) simulations and compare them with the isobaric hydrodynamic predictions. We use the dimensionless units introduced in Sec.~\ref{finite_density_form}, with $d=m=k_B=T_*=1$. The channel has dimensions $L_x\times L_y$, with $L_x\gg L_y$. We consider $N$ identical hard disks with elastic particle-particle and particle-wall collisions. The simulations are carried out with an event-driven molecular dynamics scheme, in which the system evolves from one collision event to the next~\cite{rapaport2004art}.

The initial left and right number densities are $\rho_L$ and $\rho_R$, and the mean flow velocity vanishes in both halves. We specify the common initial pressure $p_0$ and determine the temperatures from the same finite-density equation of state used in the hydrodynamic theory,
\begin{equation}
 T_L=\frac{p_0}{\rho_L Z(\rho_L)},\qquad
 T_R=\frac{p_0}{\rho_R Z(\rho_R)}.
\end{equation}
The particle velocities in each half are then sampled from Maxwell distributions with the corresponding temperatures.

For comparison with the one-dimensional theory, the channel is divided into longitudinal bins extending across the transverse direction. In each bin we measure the number density $\rho(x,t)$ and the local mean velocity $\mathbf u(x,t)$; the longitudinal hydrodynamic velocity used below is $v(x,t)=u_x(x,t)$. The local temperature is defined by
\begin{equation}
 T(x,t)=\frac{1}{2}\left\langle\left|\mathbf v_i-\mathbf u(x,t)\right|^2\right\rangle,
\end{equation}
where the average is over particles in the bin. The local pressure is computed from the virial expression~\cite{isobe2016hard}
\begin{align}
 p = \rho T - \frac{\rho}{2\, N' \Delta t'} \sum_{\text{collisions}} \vec{r}_{ij} \cdot \vec{v}_{ij},\label{pressureEDMD}
\end{align}
where $\vec{r}_{ij}=\vec{r}_i-\vec{r}_j$ is the separation vector at collision, $\vec{v}_{ij}$ is the relative velocity immediately before collision, $\Delta t'$ is the averaging interval, and $N'$ is the average number of particles in the bin.

\begin{figure*}[t]
\centering
\includegraphics[width=0.8\textwidth]{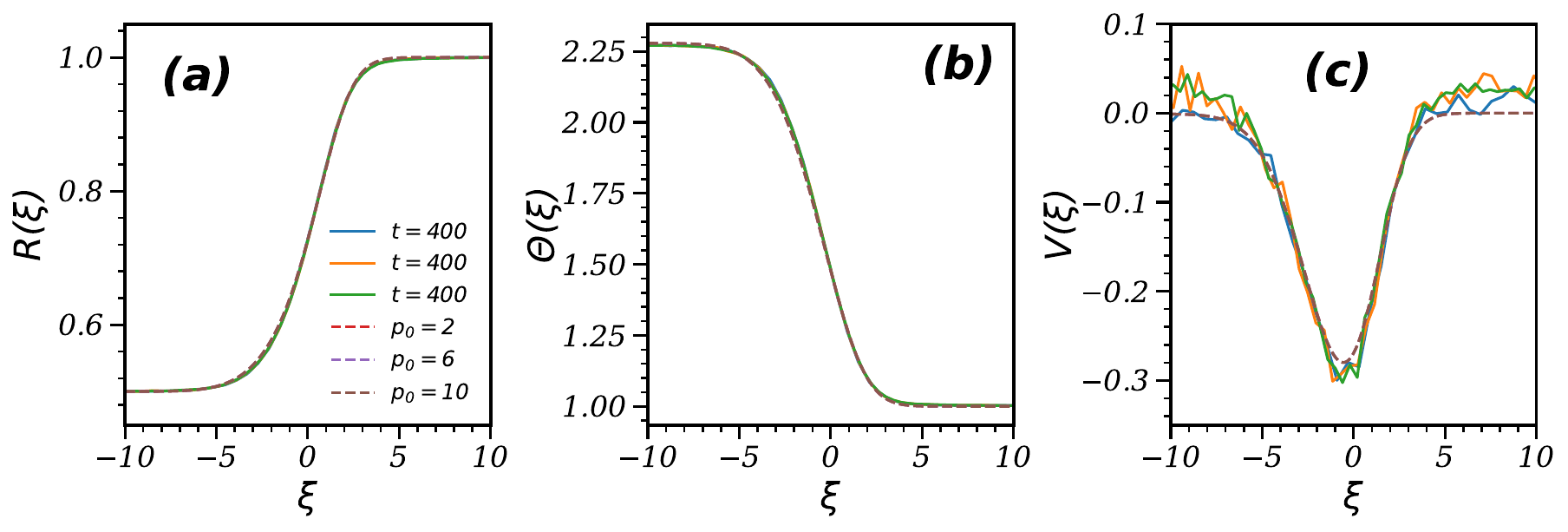}
\caption{Independence of the scaling functions from the pressure $p_0$. Shown are (a) $R$, (b) $\Theta$, and (c) $V$ as functions of $\xi$ for $p_0=2$, $6$, and $10$, with $\rho_L=0.075$ and $\rho_R=0.15$. The dashed lines are solutions of the isobaric finite-density similarity equations; the solid lines are EDMD data at $t=400$. For each pressure, $T_R=p_0/[\rho_R Z(\rho_R)]$ and $T_L=p_0/[\rho_L Z(\rho_L)]$, and the corresponding value of $D$ from Eq.~\eqref{eq:Dtilde}, with $C_1=1.418$, is used to form the similarity coordinate $\xi$. The rescaled profiles collapse onto the same curves.}
\label{NSE_all_virial_0.075_0.15_P0}
\end{figure*}

We first test the predicted independence of the scaling functions from the common pressure. Figure~\ref{NSE_all_virial_0.075_0.15_P0} compares the EDMD data with the similarity solutions for $p_0=2$, $6$, and $10$, at fixed $\rho_L=0.075$ and $\rho_R=0.15$. For each pressure, the corresponding $T_L$, $T_R$, and $D$ are used to form the similarity coordinate $\xi$. The rescaled EDMD profiles collapse onto the same hydrodynamic curves, in agreement with the prediction of Sec.~\ref{HDsolutions}.

\begin{figure*}[t]
\centering
\includegraphics[width=0.8\textwidth]{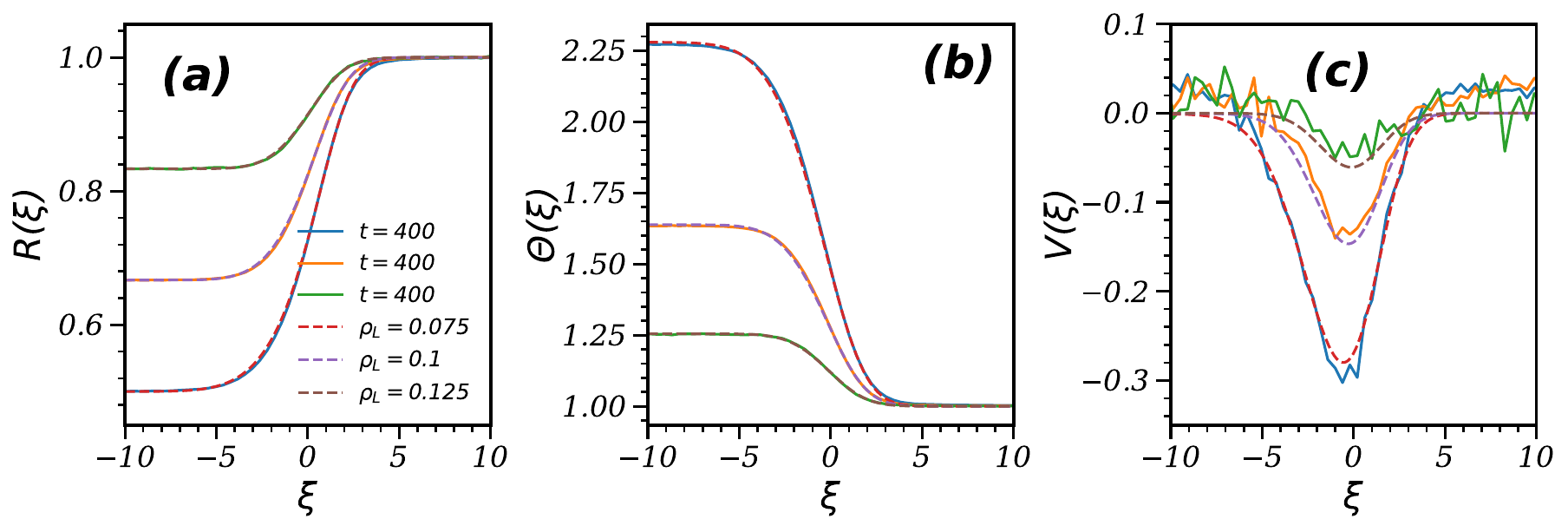}
\caption{Dependence of the finite-density solution on the density ratio $\rho_L/\rho_R$. Shown are (a) $R$, (b) $\Theta$, and (c) $V$ as functions of $\xi$, obtained from the similarity equations and from the EDMD simulations at $t=400$ for $\rho_L=0.075$, $0.1$, and $0.125$, with $\rho_R=0.15$ fixed. The solid lines are EDMD data, and the dashed lines are the solutions of the isobaric finite-density similarity equations}. The remaining parameters are $p_0=10$, $T_R=p_0/[\rho_R Z(\rho_R)]$, and $T_L=p_0/[\rho_L Z(\rho_L)]$. The hydrodynamic curves use $C_1=1.418$.
\label{NSE_all_virial_0.075_0.15_rhoR_rhoL}
\end{figure*}

We next vary the density ratio. Figure~\ref{NSE_all_virial_0.075_0.15_rhoR_rhoL} shows the EDMD data and the similarity solutions for $\rho_L=0.075$, $0.1$, and $0.125$, with $\rho_R=0.15$ and $p_0=10$ fixed. The agreement remains very good throughout this range. As $\rho_L$ increases, the density and temperature contrasts decrease, while the velocity profile changes in both amplitude and position.

\begin{figure*}[t]
\centering
\includegraphics[width=0.8\textwidth]{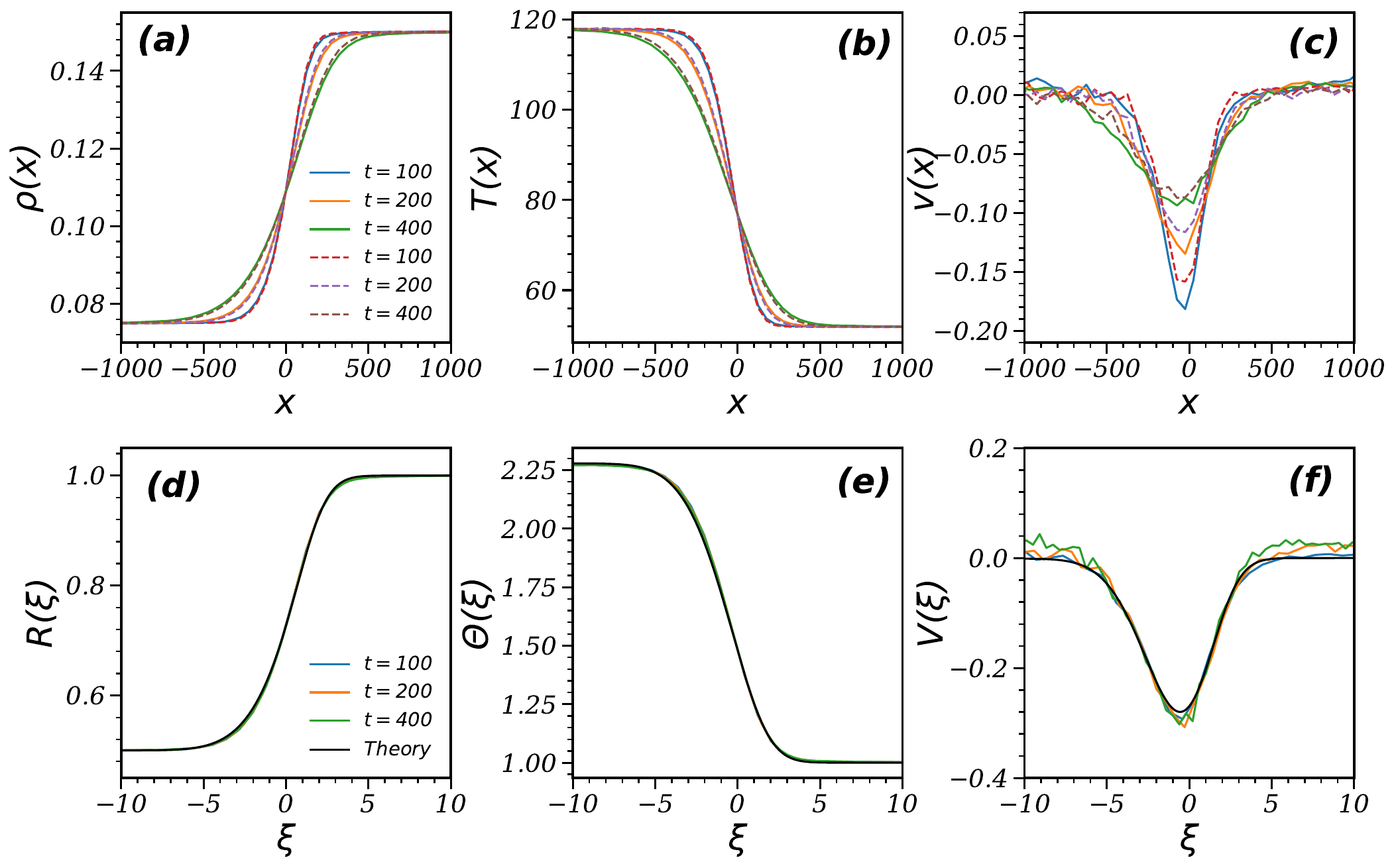}
\caption{Spatio-temporal behavior of (a--c) density, temperature, and longitudinal velocity as functions of $x$ at $t=100$, $200$, and $400$, and (d--f) the corresponding scaling functions and comparison with the hydrodynamic predictions. In panels (a--c), the solid colored lines represent EDMD data for $L_y=10$ and the dashed colored lines represent data for $L_y=20$. In panels (d--f), the EDMD data for $L_y=10$ are shown together with the solutions of the isobaric similarity equations. The results are for $p_0=10$, $\rho_R=0.15$, $\rho_L=0.075$, $T_R=p_0/[\rho_R Z(\rho_R)]$, and $T_L=p_0/[\rho_L Z(\rho_L)]$. The typical simulation box has $L_x=10^4$, with $N=11{,}250$ hard disks for $L_y=10$ and $N=22{,}500$ for $L_y=20$. The data are averaged over $2.1\times10^5$ initial configurations. The corresponding speeds of sound are approximately $17.3$ on the left and $13$ on the right. The hydrodynamic curves use $C_1=1.418$ in Eq.~\eqref{eq:Dtilde}.}
\label{all_virial_0.075_0.15_temporal}
\end{figure*}

We now turn to the time-dependent profiles and their scaling collapse. Unless stated otherwise, we use $\rho_R=0.15$, $\rho_L=0.075$, $p_0=10$, and $L_y=10$. Figures~\ref{all_virial_0.075_0.15_temporal}(a--c) show the density, temperature, and longitudinal velocity as functions of $x$ at $t=100$, $200$, and $400$, for $L_y=10$ and $20$. The transition region broadens with time, while the profiles depend only weakly on the channel width. Figures~\ref{all_virial_0.075_0.15_temporal}(d--f) show the same EDMD data in similarity variables. The data at the three times collapse well and agree closely with the solutions of the isobaric similarity equations.

The scaling functions themselves are independent of $C_1$, but comparison with EDMD data at specified $x$ and $t$ requires the conversion $\xi=x/\sqrt{Dt}$, and hence depends on $C_1$ through $D$. We obtain the best agreement for $C_1=1.418$, and use this value throughout the EDMD comparisons unless stated otherwise. For comparison, if the dilute first-Sonine form $\kappa=2C_1\sqrt{T/\pi}$ is used instead of the finite-density expression~\eqref{kappaDef}, the best-fit value is $C_1=1.595$.

\begin{figure}
\centering
\includegraphics[width=0.35\textwidth]{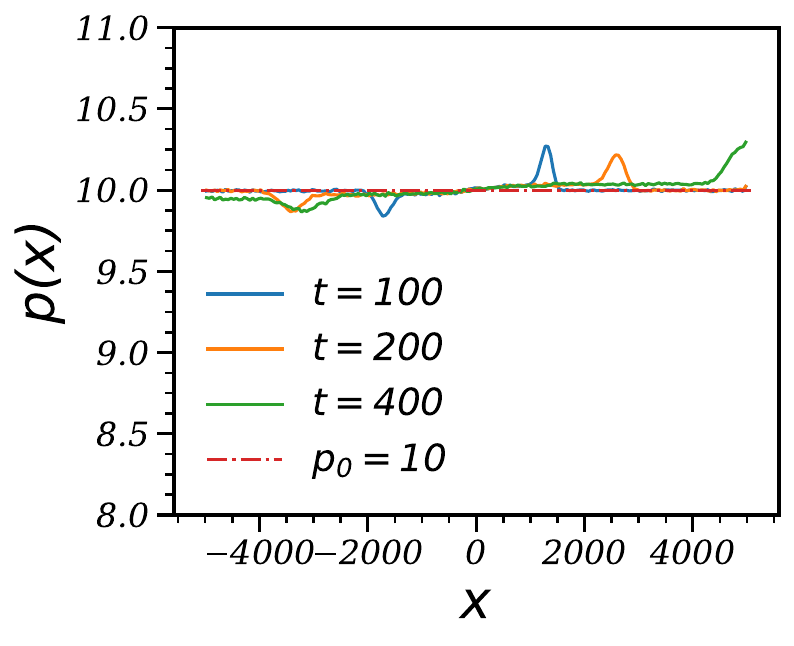}
\caption{Pressure profiles measured in the EDMD simulations at $t=100$, $200$, and $400$. The dashed horizontal line indicates the initial pressure $p_0=10$. The parameters are $\rho_R=0.15$, $\rho_L=0.075$, $T_R=p_0/[\rho_R Z(\rho_R)]$, and $T_L=p_0/[\rho_L Z(\rho_L)]$. The channel dimensions are $L_x=10^4$ and $L_y=10$, with $N=11{,}250$ hard disks. The data are averaged over $2.1\times10^5$ initial configurations. The corresponding speeds of sound are approximately $17.3$ on the left and $13$ on the right.}
\label{pressure_field_0.075_0.15_temporal}
\end{figure}

Finally, we verify directly the isobaric assumption. Figure~\ref{pressure_field_0.075_0.15_temporal} shows the pressure measured from the EDMD data using Eq.~\eqref{pressureEDMD}. The pressure remains close to the initial value $p_0=10$ over the region relevant to the comparison. At $t=400$ a pressure disturbance has already reached the longitudinal boundaries, but the reflected waves do not affect the central region $|x|\le 2000$; the thermal transition region itself remains within approximately $|x|\le1000$.

\section{Summary and Discussion}
\label{Discussion}

We have studied strongly nonlinear heat-driven flow in a gas of hard disks confined to a long channel, in a regime where the hydrodynamic velocity is small compared with the sound speed and the pressure remains nearly uniform. On time scales long compared with the acoustic time, the full compressible hydrodynamic description can therefore be reduced to an isobaric one, in which heat conduction is accompanied by a slow redistribution of mass.

In the dilute limit, this problem reduces to the previously known ideal-gas theory of Ref.~\cite{Meerson1989}. We extended the description to finite density by incorporating the hard-disk equation of state and finite-density thermal conductivity. The resulting equations retain the diffusive self-similarity of the dilute problem but yield scaling functions that differ appreciably from their ideal-gas counterparts already at the moderate densities considered here. Thus finite-density thermodynamic and transport effects are essential for a quantitative description of the system.

We tested these predictions directly by event-driven molecular dynamics. The simulations exhibit the predicted self-similar broadening of the transition region and collapse of the density, temperature, and velocity profiles at different times. They also reproduce the predicted independence of the scaling functions from the common pressure and their dependence on the density ratio. Direct measurements of the pressure show that it remains nearly uniform in the central region where the comparison is performed. With a single fitted overall conductivity factor, $C_1=1.418$, the isobaric hydrodynamic theory gives very good quantitative agreement with the particle simulations over all of these tests.

The fitted value of $C_1$ should be viewed as an effective transport coefficient for the finite-width channel. In two dimensions the thermal conductivity is affected by long-wavelength hydrodynamic fluctuations, while the finite transverse size $L_y$ cuts off the two-dimensional logarithmic renormalization and may introduce additional quasi-one-dimensional finite-size effects. These effects are not treated explicitly here and can be absorbed, to the accuracy relevant for the present simulations, into the overall conductivity factor.

The present results complement particle-level tests of hydrodynamics in rapidly evolving high-Mach-number flows such as shocks and blast waves. Here the opposite regime is realized: acoustic degrees of freedom relax rapidly, while heat conduction and slow mass redistribution govern the subsequent evolution. To our knowledge, the simulations provide the first direct particle-level test of zero-Mach-number hydrodynamics for a strongly inhomogeneous, thermally driven gas. The agreement obtained for density, temperature, velocity, and pressure shows that the isobaric reduction gives a quantitatively accurate description of the underlying hard-disk dynamics even when the temperature and density contrasts are large.

~

\textbf{Acknowledgments.}   We thank Andr{\'e}s Santos for a useful discussion of finite-width corrections to the hard-disk equation of state. We thank the Rutgers Mathematical Physics Webinar series, organized by Joel Lebowitz, which led to this collaboration. All the simulations were carried out on Supercomputers Contra, Tetris, and Mario at ICTS-TIFR Bengaluru. A.K. and A.D. would like to acknowledge the support from the DAE, Government of India, under Project No. RTI4001. AD acknowledges the J.C. Bose Fellowship (JCB/2022/000014) of the Science and Engineering Research Board of the Department of Science and Technology, Government of India.  B.M. was supported by the Israel Science Foundation (Grant No. 1579/25).

%

\end{document}